\documentclass[]{mn2e}
\usepackage{graphicx}

\newcommand \mum {\,{\rm \mu m}}

\newcommand \asil {a_{\rm sil}}

\newcommand \Vsil {V_{\rm sil}}
\newcommand \Vice {V_{\rm ice}}

\newcommand \water {\rm H_2O}

\newcommand \magni {\,{\rm mag}}
\newcommand \sgrA {{\rm Sgr}\,{\rm A}^{\ast}}
\newcommand \simali {{\sim\,}}
\newcommand \DeltaY {{\Delta Y}}
\newcommand \Cabsam {{C_{\rm abs}^{\rm am}}}
\newcommand \Cabscrst {{C_{\rm abs}^{\rm crys}}}
\newcommand       \simlt        {\leq}

\title[On the Crystallinity of Interstellar Silicates]
      {On the Crystallinity of Silicate Dust in the Interstellar Medium}

\author[M.~P.~Li, G.~Zhao \& A.~Li]
       {M.~P.~Li$^{1}$, G.~Zhao$^{1}$,
        and Aigen Li$^{2}$\thanks{%
                E-mail: lmp@bao.ac.cn, gzhao@bao.ac.cn,
                        lia@missouri.edu}\\
       $^{1}$National Astronomical Observatories,
             Chinese Academy of Sciences,
             Beijing 100012, China\\
       $^{2}$Department of Physics and Astronomy,
             University of Missouri,
             Columbia, MO 65211, USA}
\begin{document}
\date{Received date  / Accepted date }
\pagerange{\pageref{firstpage}--\pageref{lastpage}} \pubyear{2007}

\maketitle

\label{firstpage}
\begin{abstract}
An accurate knowledge of the mineralogy
(chemical composition and crystal structure)
of the silicate dust in the interstellar medium (ISM)
is crucial for understanding its origin in evolved stars,
the physical and chemical processing in the ISM,
and its subsequent incorporation into protostellar
nebulae, protoplanetary disks and cometary nuclei
where it is subjected to further processing.
While an appreciable fraction of silicate dust
in evolved stars, in protoplanetary disks around
pre-main sequence stars, in debris disks
around main sequence stars, and in cometary nuclei
is found to be in crystalline form,
very recent infrared spectroscopic studies
of the dust along the sightline toward the Galactic
center source $\sgrA$ placed an upper limit of
$\simali$1.1\% on the silicate crystalline fraction,
well below the previous estimates of $\simali$5\%
or $\simali$60\% derived from the observed 10$\mum$
absorption profile for the local ISM toward Cyg OB2 No.12.
Since the sightline toward $\sgrA$ contains molecular
cloud materials as revealed by the detection of
the 3.1 and 6.0$\mum$ water ice absorption features,
we argue that by taking into account
the presence of ice mantles on silicate cores,
the upper limit on the degree of silicate crystallinity
in the ISM is increased to $\simali$3--5\%.
\end{abstract}


\section{Introduction}
The mineralogical composition of dust contains
important information about its origin and evolution,
and may reveal the physical, chemical and evolutionary
properties of the astrophysical regions 
where the dust is found. Silicate, one of the major 
components of cosmic dust species, is ubiquitously 
seen in various astrophysical environments,
ranging from the Galactic diffuse interstellar medium (ISM),
HII regions, and the dust torus around active galactic
nuclei to dust envelopes around evolved stars,
protoplanetary disks around pre-main sequence stars,
debris disks around main sequence stars, 
cometary comae and interplanetary space.
Their chemical composition and crystal structure
vary with local environments.

Interstellar spectroscopy provides
the most diagnostic information on dust composition.
In the infrared (IR), the strongest interstellar spectral
features are the 9.7 and 18$\mum$ absorption (or emission)
bands which are generally attributed to
the Si--O stretching and O--Si--O bending modes, respectively.
While the absorption profiles measured in laboratory
for crystalline olivine and pyroxene show many sharp features,
in the diffuse ISM, the observed 9.7 and 18$\mum$ silicate
features are broad and relatively featureless,
suggesting that interstellar silicates
are mainly amorphous rather than crystalline.
Since the crystallinity of silicate dust is intimately connected
to the energetic processing occurred in the evolutionary
life cycle of dust and depends on the environmental properties,
of particular interest to silicate astromineralogy is
the abundance of crystalline silicates in
different astrophysical environments,
especially in the diffuse ISM.

Recently, a number of studies have been made to
determine the crystallinity degree of silicates
in the diffuse ISM.
From the observed 9.7$\mum$ absorption
profile for dust in the local ISM toward Cyg OB2 No.12,
Li \& Draine (2001) estimated the fraction
of Si in crystalline silicates to be $\simlt$5\%.
Demyk et al.\ (1999) derived an upper limit of
$\simali$1--2\% of crystalline silicates in mass
toward two massive protostars.
However, Bowey \& Adamson (2002) argued that
a complex mixture of crystalline silicates (60\% by mass)
and amorphous silicates (40\% by mass) could explain
the observed smooth silicate absorption profile
at 9.7$\mum$ toward Cyg OB2 No.12.
More recently, Kemper, Vriend \& Tielens (2004) placed
a more strict constraint of $\simlt 1.1\%$
on the crystallinity of interstellar silicates,
based on a detailed analysis of the 9.7$\mum$
feature obtained with the Short Wavelength Spectrometer (SWS)
on board the Infrared Space Observatory (ISO).
This upper limit $\simali$1.1\% of crystalline
silicates was derived from a direct comparison of
the $\sgrA$ spectrum with theoretical spectra
for pure silicates.

However, it is evident from the detection
of the 3.1 and 6.0$\mum$ water ice features
respectively attributed to the O--H stretching 
and bending modes that there are molecular cloud materials
along the line of sight toward $\sgrA$
(McFadzean et al.\ 1989; Tielens et al.\ 1996;
Chiar et al.\ 2000).
This is further confirmed by the detection of
solid CO$_2$ absorption toward $\sgrA$
(Lutz et al.\ 1996; de Graauw et al.\ 1996).
In order to infer the precise mineralogical composition 
of dust, we have to take the grain ice mantles into account
when interpreting the observed silicate features.
Indeed, in this paper we show that ignoring the ice mantles 
coated on the silicate cores can result in
an underestimation of the crystalline degree of silicate dust,
as much as $\simali$3--5\%.

  The purpose of this work is to address the question 
  how the ice mantles affect the determination of dust 
  mineralogical composition (i.e. whether and to what degree
  the inclusion of ice mantles will hide the sharp features 
  of crystalline silicates 
  and lead to an underestimation of the silicate crystallinity), 
  not to model any specific astronomical objects.
  Therefore, neither the specific choice of silicate dielectric 
  functions nor the precise constituents of the ice mantles 
  would affect our conclusion.

\section{Volume Fraction of Water Ice}
The sightline toward the Galactic center source
$\sgrA$ suffers about $\simali$30$\magni$ of
visual extinction (e.g. see McFadzean et al.\ 1989),
to which molecular clouds may contribute as much as
$\simali$10$\magni$ (Whittet et al.\ 1997).
Except for $\water$ features, there are many
additional molecular absorption features,
attributed to ${\rm CO_2}$, ${\rm NH_3}$, CO,
${\rm CH_3OH}$, ${\rm CH_4}$ and other species.
These icy molecules might accrete on the preexisting
silicate core inside dense molecular clouds
and form an ice mantle.
As for further evolution,
the accreted icy grain mantles could be
photoprocessed and converted into
organic refractory residues
(Greenberg et al.\ 1995).

Assuming the 3.1$\mum$ water ice absorption results from
the water ice mantles uniformly coated on silicate
grains which produce the 9.7$\mum$ absorption,
we estimate the volume ratio of the ice mantles
to the silicate cores from the observed optical depths
-- $\tau_{3.1}$ for the 3.1$\mum$ O--H feature,
   and $\tau_{9.7}$ for the 9.7$\mum$ Si--O feature.
Let  $\Vsil$ and $\Vice$ respectively
be the volumes of the silicate cores and the ice mantles.
Let $A_{\rm ice}^{3.1}$ and $A_{\rm sil}^{9.7}$
be the integrated band strength for
the 3.1$\mum$ O--H feature
and the 9.7$\mum$ Si--O feature, respectively.
For spherical silicate core-ice mantle grains,
we use Mie theory (Bohren \& Huffman 1983) to obtain
$A_{\rm ice}^{3.1} = \int_{3.1\mum}
(C_{\rm ice}^{3.1}/V) d\lambda$
and
$A_{\rm sil}^{9.7} = \int_{9.7\mum}
(C_{\rm sil}^{9.7}/V) d\lambda$,
where $C_{\rm ice}^{3.1}$ ($C_{\rm sil}^{9.7}$)
is the continuum-subtracted absorption cross section
of the 3.1$\mum$ (9.7$\mum$) ice (silicate) feature,
and $V$ is the volume of a spherical grain with radius $a$.
For the $\sgrA$ sightline,
$\tau_{3.1}\approx 0.50$ and
$\tau_{9.7}\approx 3.52$
(Chiar et al.\ 2000).
We therefore obtain
$\Vice/\Vsil=\left(\tau_{3.1}/\tau_{9.7}\right)
\left(A_{\rm sil}^{9.7}/A_{\rm ice}^{3.1}\right)
\approx0.55$, 
assuming that all silicate grains 
are coated by a layer of ice mantle.\footnote{%
  In reality, a thicker 
  ice mantle would be expected for the silicate
  dust in the molecular cloud component
  which contributes about 10$\magni$ to
  the total $\simali$30$\magni$ of
  visual extinction toward $\sgrA$.
  An ice mantle as thick as
  $\Vice/\Vsil\approx 7.6$
  was implied for the dust
  in the dense molecular cloud
  toward the field star Taurus Elias 16  
  (Bowey et al.\ 1998).
  A thicker ice mantle would hide up a larger
  fraction of crystalline silicates.
  }

\section{Spherical Grains}
We consider 3 types of dust materials: amorphous
olivine, crystalline olivine, and pure water ice. We adopt the
dielectric functions of Dorschner et al.\ (1995) for amorphous
olivine, of Li \& Draine (2001) for crystalline olivine, and of
Hudgins et al.\ (1993) for water ice.

We first consider spherical
silicate core-water ice mantle grains.
We take the silicate core size to be $\asil$\,=\,0.1$\mum$,
a typical size for interstellar dust.
We should note that the choice of an exact
grain size is not critical since in the wavelength
range considered here submicron-sized
interstellar grains are in the Rayleigh regime.
Also because of this, we do not need to consider
dust size distributions.

\begin{figure}
\centering
\includegraphics[width=8cm]{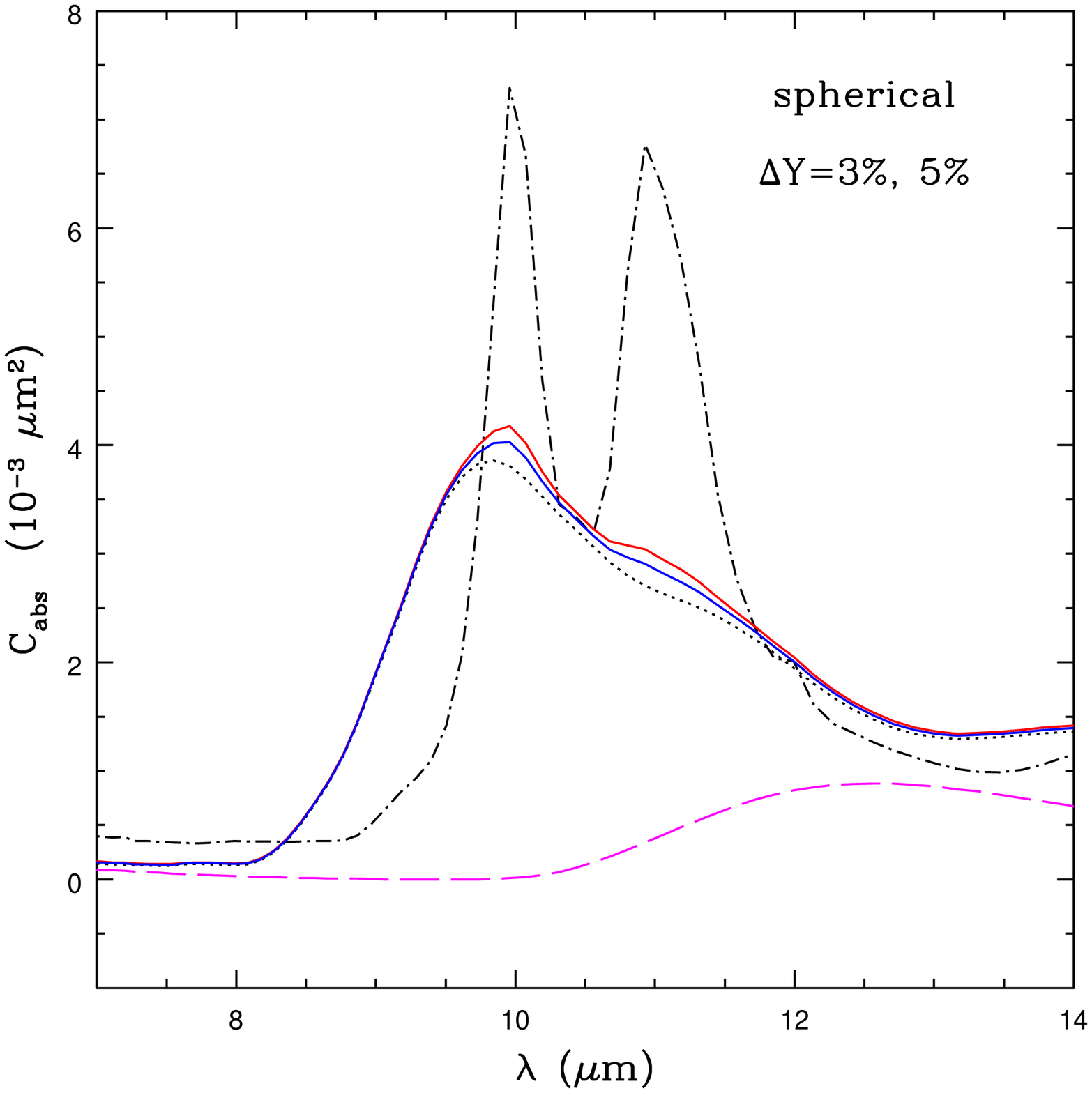}
\caption{ 
         \label{fig:sphere} 
         Absorption cross sections of spherical grains 
         for $\Vice/\Vsil=0.55$. Dotted: amorphous silicate 
         core-ice mantle grains ($\Cabsam$);
         dot-dashed: crystalline olivine silicate core-ice mantle 
         grains ($\Cabscrst$);
         long-dashed: equal-volume ice spheres 
         for the ice mantles; 
         solid: $C_{\rm abs} = \Cabsam + \DeltaY \Cabscrst$ 
         where $\DeltaY$\,=\,5\% is the crystalline olivine 
         silicate fraction; 
         short-dashed: 
         $C_{\rm abs} = \Cabsam + \DeltaY \Cabscrst$ 
         with $\DeltaY$\,=\,3\%.
         Note that water ice produces a shoulder 
         at $\lambda > 11\mum$. This should not
         be interpreted as a crystalline silicate feature.
         }
\end{figure}

In Figure \ref{fig:sphere}
we plot the absorption cross sections for
amorphous olivine silicate core-$\water$ ice mantle grains and 
for crystalline olivine silicate core-$\water$ ice mantle grains.
As shown in Figure \ref{fig:sphere},
the absence of narrow features
near 10.0 and 11.1$\mum$ in the absorption spectrum
suggests that the inclusion of water ice mantles
would hide up $\simali$3--5\% crystalline mass fraction
without being noticed. 
Note that water ice has a broad absorption band 
at $\simali$12.2$\mum$ 
(see Fig.\,\ref{fig:sphere}; 
L\'eger et al.\ 1983; Hudgins et al.\ 1993).
The inclusion of an ice mantle on silicate dust
(no matter it is amorphous silicate or crystalline silicate)
results in a weak shoulder at $\lambda > 11\mum$.
This is noticeable even in the model spectrum of
ice-coated pure amorphous silicate dust
(see Fig.\,\ref{fig:sphere}).
One should caution that this should not be interpreted 
as a crystalline silicate feature.

\begin{figure}
\centering
\includegraphics[width=8cm]{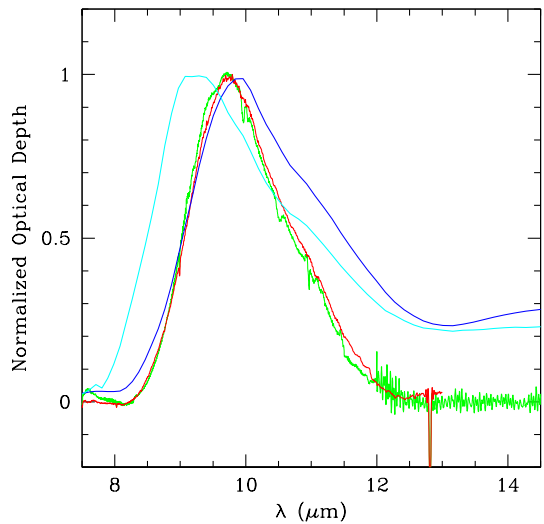}
\caption{ 
         \label{fig:real} 
         Comparison of the ISO spectra of $\sgrA$
         (thick solid: Gibb et al.\ 2004;
          thick dashed: Kemper et al.\ 2004)
         and the absorption spectra of 
         olivine (thin solid) 
         and pyroxene (dot-long dashed) spheres
         with 3\% crystallinity, with 1/3 of the dust
         coated by a layer of ice mantle
         ($\Vice/\Vsil = 0.83$) and 2/3 of the dust
         being bare silicates.
         The $\lambda>11\mum$ shoulder of 
         the model spectra arises from the ice mantles 
         (also see Fig.\,\ref{fig:sphere}).
         Also shown is the absorption spectrum of
         pure amorphous olivine dust without an 
         ice mantle (dot-short dashed).
        }
\end{figure}

We also consider a model in which only the molecular cloud 
dust is coated with a layer of ice mantle. If we assume that 
silicate dust and carbon dust equally contribute to 
the $\simali$10\,mag visual extinction, the volume ratio
of the ice mantle to the silicate core for the molecular
cloud dust would be $\Vice/\Vsil \approx 0.83$. 
The bare diffuse cloud dust is responsible for 
the remaining $\simali$20\,mag visual extinction.  
In Figure \ref{fig:real} we show the model cross sections
obtained by summing up 1/3 of that produced by
the molecular cloud dust and 2/3 of that produced by
the diffuse cloud dust. Similarly, $\simali$3--5\%
crystalline silicate would be hidden because of 
the presence of an ice mantle on the molecular 
cloud dust toward $\sgrA$.\footnote{%
  Admittedly, the approach adopted here is simplified.
  The molecular cloud silicate dust may differ from
  that of the diffuse cloud (e.g. their 10$\mum$
  Si--O features may have different line profiles 
  [particularly line widths], 
  see Bowey, Adamson, \& Whittet 2001 
  and references therein).
  Their optical properties which are temperature-dependent
  (Bowey et al.\ 2001) may also differ from each other 
  since the molecular cloud dust is generally
  colder than the diffuse cloud dust
  (e.g. see Greenberg \& Li 1996a).
  Moreover, it is not clear if the molecular cloud
  silicate dust has the same iron fraction
  as the diffuse cloud silicate dust
  (it is well recognized that, as most recently 
   experimentally demonstrated by Bowey et al.\ [2007],
   the optical properties of silicate dust
   are very sensitive to the proportion of iron).
  }

So far, we have only considered olivine silicate dust.
It is possible that there is a considerable
amount of pyroxene dust in the ISM 
(see Bowey \& Adamson 2002 and references therein).
For illustration, we carry out similar calculations
for pyroxene and obtain similar conclusions
(see Fig.\,\ref{fig:real}).  
Again, we should stress that the purpose of this work 
is not to model any specific astronomical objects,
but to investigate the effects of ice mantles
on the silicate crystallinity estimation.
We do not expect either olivine or pyroxene
alone could closely reproduce the interstellar
silicate absorption feature observed toward $\sgrA$.
For comparison, we also show in Figure\,\ref{fig:real} 
the ISO spectra of $\sgrA$ obtained by
Kemper et al.\ (2004) and Gibb et al.\ (2004).\footnote{%
  We note that the ISO spectrum of $\sgrA$
  appears to have a feature at $\simali$6.9$\mum$,
  attributed to crystalline melilite
  (Bowey \& Hofmeister 2005).
  }
The fact that, while the 10$\mum$ feature of olivine 
peaks at relatively longer wavelengths than the observed
spectra, pyroxene peaks at too short wavelengths,
suggests the co-existence of both olivine and
pyroxene in the ISM.

\section{Spheroidal Grains}
However, interstellar grains must be nonspherical
as indicated by interstellar polarization.
We now consider spheroidal shape which can be exactly
solved in the Rayleigh limit
(see Li et al.\ 2002 and references therein).

We consider confocal silicate core-ice mantle spheroidal
grains with a volume-equivalent sphere radius
$r_{\rm sil}^{\rm eq}={(ab^2)}^{1/3}=0.1\mum$,
where $a$ and $b$ are the semi-axis along and
perpendicular to the symmetry axis, respectively.
Again, we take the volume ratio of the ice mantles
to the silicate cores to be $\Vice/\Vsil = 0.55$.

\begin{figure}
\centering
\includegraphics[width=9cm]{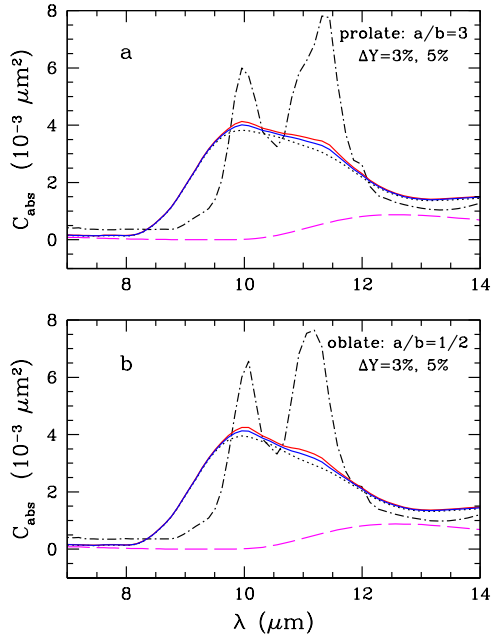}
\caption{
\label{fig:spheroid}
Same as Figure \ref{fig:sphere}
but for (a) $a/b=3$ prolates
and (b) $a/b=1/2$ oblates.
}
\end{figure}

We first consider prolate grains with $r_{\rm sil}^{\rm
eq}=0.1\mum$ for the silicate cores
and $a/b=3$ for the ice mantles.\footnote{%
  Greenberg \& Li (1996b) found that prolates of $a/b=3$
  provide an almost prefect match to
  the 10, 18$\mum$ silicate polarization features
  of the Becklin-Neugebauer (BN) object.
  }
In Figure \ref{fig:spheroid}a we plot the absorption
cross sections for ice coated-amorphous silicate
prolates and ice coated-crystalline prolates.
Just like spherical grains, the inclusion of
water ice mantles would hide up $\simali$3\%
crystalline mass fraction without being noticed.
Similarly, we perform the same calculations but for
oblate grains with $a/b=1/2$ for ice mantles.\footnote{%
  Lee \& Draine (1985) and Hildebrand \& Dragovan (1995)
  found that $a/b=1/2$ oblates fit well the 3.1$\mum$ ice
  polarization and the 10$\mum$ silicate polarization of
  the BN object.
  }
As seen in Figure \ref{fig:spheroid}b,
a crystallinity degree of as much as
$\simali$3\% can be covered by the ice mantles.

\section{Grains with Distributions of Ellipsoidal Shapes}
The shape of a spheroidal grain is characterized
by its eccentricity $e$
(and the depolarization factors
$L^{\parallel}$, $L^{\bot}$ via eq.[4] of Li et al.\ 2002).
In \S4 we consider grains with a single shape.
We now consider an average of a distribution of shapes.
We assume two kinds of shape distribution functions:
1) $dP/dL^{\parallel}$ = constant, i.e.,
   all shapes are equally probable
   (Bohren \& Huffman 1983);
2) $dP/dL^{\parallel} = 12 L^{\parallel}[1-L^{\parallel}]^2$
   (Ossenkopf, Henning, \& Mathis 1992).\footnote{%
   This distribution peaks at spheres ($L^{\|}=L^{\bot}$=1/3)
   and is symmetric about spheres with respect to
   eccentricity $e$. It drops to zero for
   the extreme cases:
   for infinitely thin needles ($a\gg b$)
   $L^{\parallel}\rightarrow 0$;
   for infinitely flattened pancakes ($a\ll b$)
   $L^{\parallel}\rightarrow 1$.
   }

Averaging over the shape distribution,
we have the resultant absorption cross section
$C_{\rm abs} = \int_{0}^{1} dL^{\parallel} dP/dL^{\parallel}
C_{\rm abs}(L^{\parallel})$,
where $C_{\rm abs}(L^{\parallel})$ is the absorption
cross section of a particular shape $L^{\parallel}$
(note $L^{\parallel}$ is for the mantle;
the core depolarization factor is derived
from eqs.[4-6] of Li et al.\ 2002).
Again, we assume confocal geometry for core-mantle grains,
with the above $dP/dL^{\parallel}$ (as well as $e$) applying
to the outer surface.

\begin{figure}
\centering
\includegraphics[width=9cm]{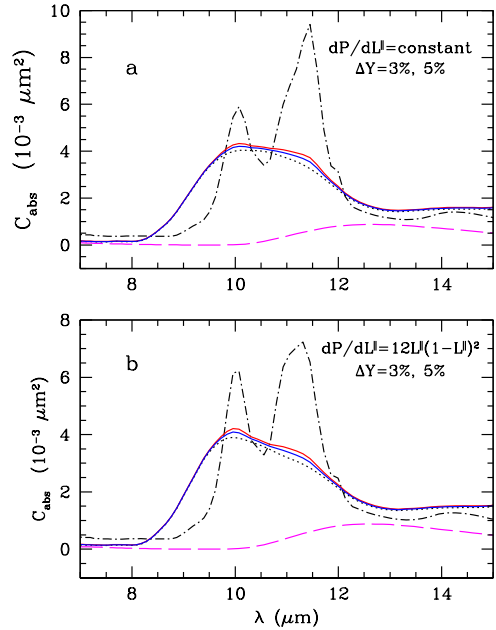}
\caption{
\label{fig:cde}
Same as Figure \ref{fig:sphere} but
(a) for grains with a uniform distribution
    of ellipsoidal shapes
    (i.e. $dP/dL^{\parallel}$=constant)
    and
(b) for grains with
    $dP/dL^{\parallel} = 12 L^{\parallel}[1-L^{\parallel}]^2$.
}
\end{figure}

The results are shown in Figure \ref{fig:cde}.
Similar to spherical grains and grains of a single
ellipsoidal shape, with $\simali$3\% crystalline silicates
included, the absorption profiles still do not seem to
exhibit the sharp features of crystalline silicates.

\section{Summary}
We have investigated the effects of ice mantles
coated on silicate cores of various shapes
on the determination of the crystallinity
degree of silicates. It is found that for
the dust in the line of sight toward the Galactic
center source $\sgrA$, $\simali$3\% crystalline
silicates could be hidden by ice mantles,
well exceeding the upper limit
$\simali$1.1\% of Kemper et al.\ (2004)
derived from the assumption of the absence of
dense molecular materials in this line of sight.
We take the standard approach by assuming that
the silicates are amorphous and then adding in
crystalline components.
But as shown in Kemper et al.\ (2004),
the smooth, featureless 10$\mum$ amorphous silicate
spectrum allows as much as $\simali$2.2\%
crystalline silicates (or even higher; see Bowey \& Adamson 2002).
Therefore, the total allowable degree of
crystallinity would be $\simali$5\%,
consistent with the earlier estimates
of Li \& Draine (2001).

\section*{Acknowledgments}
We thank J.Y. Hu and S.L. Liang for helpful comments.
We thank the referee whose report improved 
our paper significantly. 
We thank J.E. Bowey and F. Kemper 
for providing us with the ISO spectra of $\sgrA$.
ML and GZ are supported by
the NSFC Grants 10433010 and 10521001.
AL is supported in part
by the University of Missouri Research Board,
a NASA/HST Theory Program grant,
a NASA/Spitzer Theory Program grant,
and the NSFC Outstanding Oversea Young Scholarship.

\bsp
\label{lastpage}


\begin{thebibliography}{}
%
\bibitem[]{}Bohren, C. F., \& Huffman, D. R.\ 1983,
            Absorption and Scattering of Light by
            Small Particles (New York: Wiley)
\bibitem[]{}Bowey, J. E., \& Adamson, A. J.\
            2002, MNRAS, 334,94
\bibitem[]{}Bowey, J.~E., \& Hofmeister, A.~M.\ 2005, 
            MNRAS, 358, 1383 
\bibitem[]{}Bowey, J.~E., Adamson, A.~J., \& Whittet, D.~C.~B.\ 
            1998, MNRAS, 298, 131 
\bibitem[]{}Bowey, J.~E., Adamson, A.~J., \& Yates, J.~A.\ 2003,
            MNRAS, 340, 1173
\bibitem[]{}Bowey, J.~E., Rawlings, M.~G., \& Adamson, A.~J.\
            2004, MNRAS, 348, L13
\bibitem[]{}Bowey, J.~E., Morlok, A., K{\"o}hler, M., 
            \& Grady, M.\ 2007, MNRAS, 376, 1367
\bibitem[]{}Bowey, J.~E., Lee, C., Tucker, C., Hofmeister, A.~M., 
            Ade, P.~A.~R., \& Barlow, M.~J.\ 
            2001, MNRAS, 325, 886 
\bibitem[]{}Chiar, J. E., Tielens, A. G. G. M., Whittet, D. C. B.,
            Schutte, W. A., Boogert, A. C. A., Lutz, D.,
            Van Dishoeck, E. F., \& Bernstein, M. P.\
            2000, ApJ, 537, 749
\bibitem[]{}de Graauw, Th., et al.\ 1996, A\&A, 315, L345
\bibitem[]{}Demyk, K., Jones, A. P., Dartois, E., Cox, P.,
            \& d'Hendecourt, L.\ 1999, A\&A, 349, 267
\bibitem[]{}Dorschner, J., Begemann, B., Henning, Th.,
            J\"ager, C., \& Mutschke, H. 1995, A\&A, 300, 503
\bibitem[]{}Gibb, E.L., Whittet, D.C.B., Boogert, A.C.A., 
            \& Tielens, A.G.G.M.\ 2004, ApJS, 151, 35 
\bibitem[]{}Greenberg, J. M., \& Li, A.\ 1996a, 
            New Extragalactic Perspectives in the New South Africa, 
            ed. D. L. Block \& J. M. Greenberg 
            (Dordrecht: Kluwer), 118
\bibitem[]{}Greenberg, J. M., \& Li, A. 1996b, A\&A, 309, 258
\bibitem[]{}Greenberg, J.M., Li, A., Mendoza-G\'{o}mez, C.X.,
            Schutte, W.A., Gerakines, P.A., \& de Groot, M.\
            1995, ApJ, 455, L177
\bibitem[]{}Hildebrand, R. H., \& Dragovan, M.\
            1995, ApJ, 450, 663
\bibitem[]{}Hudgins, D. M., Sandford, S. A., Allamandola, L. J.,
            \& Tielens, A. G. G. M. 1993, ApJS, 86, 713
\bibitem[]{}Kemper, F., Vriend, W. J., \& Tielens, A.G.G.M.\
            2004, ApJ, 609, 826 (erratum: 2005, ApJ, 633, 534)
\bibitem[]{}Lee, H. M., \& Draine, B. T.\ 1985, ApJ, 290, 211
\bibitem[]{}L\'eger, A., Gauthier, S., Defourneau, D., 
            \& Rouan, D.\ 1983, A\&A, 117, 164 
\bibitem[]{}Li, A., \& Draine, B.T. 2001, ApJ, 550, L213
\bibitem[]{}Li, A., \& Greenberg, J. M. 2002, ApJ, 577, 789
\bibitem[]{}Li, A., Greenberg, J. M., \& Zhao, G.\
            2002, MNRAS, 334, 840
\bibitem[]{}Lutz, D., et al.\ 1996, A\&A, 315, L269
\bibitem[]{}McFadzean, A. D., Whittet, D. C. B., Longmore, A. J.,
            Bode, M. F., \& Adamson, A. J. 1989, MNRAS, 241, 873
\bibitem[]{}Ossenkopf, V., Henning, Th., \& Mathis, J.S.\
            1992, A\&A, 261, 567
\bibitem[]{}Tielens, A. G. G. M., Wooden, D. H., Allamandola, L. J.,
            Bregman, J., \& Witteborn, F. C. 1996, ApJ, 461, 210
\bibitem[]{}Whittet, D. C. B., et al.\ 1997, ApJ, 490, 729
\end{thebibliography}
\end{document}